\documentclass{article}

\usepackage{PRIMEarxiv}

\usepackage[utf8]{inputenc} % allow utf-8 input
\usepackage[T1]{fontenc}    % use 8-bit T1 fonts
\usepackage{hyperref}       % hyperlinks
\usepackage{url}            % simple URL typesetting
\usepackage{booktabs}       % professional-quality tables
\usepackage{amsfonts}       % blackboard math symbols
\usepackage{nicefrac}       % compact symbols for 1/2, etc.
\usepackage{microtype}      % microtypography
\usepackage{lipsum}
\usepackage{fancyhdr}       % header
\usepackage{graphicx}       % graphics
\graphicspath{{media/}}     % organize your images and other figures under media/ folder
\usepackage{subcaption}
\usepackage{physics}
\usepackage{cuted}
\usepackage{textcomp}

\title{Fast Metasurface Hybrid Lens Design using a Semi-Analytical Model}
\author{
  A. Cléroux Cuillerier$^*$, J. Borne and S. Thibault \\
  Centre d'optique, photonique et laser, Université Laval\\
  Université Laval \\
  Quebec City, Canada\\
  $^*$ alexandre.cleroux-cuillerier.1@ulaval.ca \\
}

\begin{document}
\maketitle

\begin{abstract}
We propose a new method for integrating metasurfaces in optical design using semi-analytical modelling of dielectric nanostructures. The latter computes the output phase of an electric field incident on the metasurface, allowing their use with ray-tracing software. This tool provides a method to use metasurfaces in optical systems while using built-in optimization processes to avoid time-consuming computation. To demonstrate the applicability and versatility of our method, we present variations of a triplet composed of refractive elements and a metasurface. For each of the systems, similar optical performances are achieved. Our unique and innovative approach to joining metasurfaces and ray tracing has the potential to promote the design of innovative systems by exploiting the richness of metasurfaces and the functionality of conventional lens design software.
\end{abstract}

% keywords can be removed
\keywords{Optics \and Metasurface \and Lens Design}

\section{Introduction}

Research on metasurfaces has been motivated by their broad potential for manipulating light. Exotic applications exploiting their increased control on the wavefront, their polarization sensitivity or their unique chromatic dispersion were proposed in the past years \cite{Chen2016, Achouri2018, Chang2018, Liang2019, Hu2021}. Many of these technologies were demonstrated by wave optics computation, sometimes followed by experimental work. Indeed, high nonlinear responses \cite{Krasnok2018}, multifunctional metalens \cite{Shi2018,Holsteen2019} or non-Hermitian metasurfaces \cite{Dong2020} can be found in the literature, to cite a few.

The utility of metasurfaces has been praised many times; some even say they have the potential to replace traditional optics \cite{MOON2020101877, Tseng2018, Khorasaninejad2017}. However, their integration in conventional lens design remains a little-explored avenue. Considering metasurfaces in a complete optical system is inconceivable if common time-consuming numerical methods are used. The macroscopic scale of elements and the large number of iterations required to design a suitable system \cite{Smith1966} would require too much computing power. Therefore, a method to efficiently represent metasurfaces in a ray-tracing process is needed. Such a method could originate from modelling the optical properties of meta-atoms. 

Thus, we propose an approach for using metasurfaces in ray-tracing applications thanks to the efficient characterization of their optical properties provided by a model. Section \ref{sec:ray} describes the ray-tracing equations adapted for propagation through metasurfaces. Such equations will later be implemented in ray-tracing software like Zemax$^\textrm{\textregistered}$. Then, in section \ref{sec:theory}, a semi-analytical model computing the optical properties of a metasurface composed of dielectric nanocylinders is given. This modelling is a continuation of our previous work, where rectangular structures were studied \cite{Borne2021, Cuillerier:21}. Results obtained by this model are then compared to FDTD simulations in section \ref{sec:meta_proper}.
Finally, we present optical designs using different refractive lenses and metasurfaces targeting the same application to display the method's versatility in section \ref{sec:design+perfo}.

\section{Ray-tracing through a metasurface} \label{sec:ray}

As the ray-tracing method is based upon calculating the angle of refraction at each optical surface, it is natural to start our implementation with the generalized law of refraction for metasurfaces proposed by Yu et al. \cite{Yu:GeneralizedSnell}. Here is the case for the general refraction
\begin{align}
    n_t\sin\theta_t - n_i\sin\theta_i = \frac{\lambda}{2\pi}\pdv{\varphi}{r} , \label{eq:general_sdl}
\end{align}
where the metalens implements a phase $\varphi$ on the wavefront, and $n_i$, $\theta_i$ and $n_t$, $\theta_t$ are the refractive index and the incidence angle before and after the metasurface.

In optical design software such as Zemax$^\textrm{\textregistered}$ or CodeV$^\textrm{\textregistered}$, rays are represented by a three-dimensional vector $\vb{R} = x \vu{x} + z \vu{z} + z \vu{z}$. The orientation of such a ray is defined by its direction cosines $l$, $m$ and $n$.
\begin{align}
    l = \cos{u_x}, \quad m = \cos{u_y}, \quad n = \cos{u_z} ,
\end{align}
where $u_x$, $u_y$ et $u_z$ are the angles formed by the vector $\vb{R}$ and the axes $\vu{x}$, $\vu{y}$ and $\vu{z}$. Considering metasurfaces are flat components placed perpendicular to the optical axis in most cases, we can establish that $\theta_i = 90^\circ - u \implies \sin{\theta_i} = \cos{u}$. Thus, by rearranging Eq. \ref{eq:general_sdl} with $n_t = n_i = 1$ using the direction cosines, the following set of equations for the coordinates $(l', m', n')$ of the refracted beam can be obtained 
\begin{align}
    l' &= l + \frac{\lambda}{2\pi} \pdv{\varphi(x,y)}{x} \label{eq:cosine_l},\\
    m' &= m + \frac{\lambda}{2\pi} \pdv{\varphi(x,y)}{y} \label{eq:cosine_m},\\
    n' &= \sqrt{1 - (l')^2 - (m')^2} \label{eq:cosine_n} .
\end{align}
The normalization condition $l^2 + m^2 + n^2 = 1$ was used for the $n'$-coordinate. If one wishes to perform ray tracing through a metasurface, one must know the phase shift induced for a given incident beam

\section{Semi-analytical model}
\label{sec:theory}

As stated previously, obtaining the phase shift implemented by the metasurface via conventional full-wave simulations is inconceivable for lens design purposes considering the standard optimization process. Therefore, a model that can efficiently output the optical properties used for ray-tracing is advisable. Therefore, this section concerns the development of an expression for the phase shift of cylindrical nanostructures, an essential parameter for performing the ray tracing as previously determined.

Related to our earlier work about nanofins \cite{Borne2021}, the guided propagation through a nanocylindrical waveguide can be described as a phase accumulation of the electrical field given by the optical path of an equivalent wave in a medium of length $H$ as a first-order approximation. 
\begin{align}
    \varphi = \arg \qty{ \sum_j \tau_j \exp( -i \beta_j H ) } \label{eq:phase_general} ,
\end{align}
where $\tau_j$ and $\beta_j$ are respectively the complex amplitude and the propagation constant of the waveguide mode $j$. For the needs of our application, we consider the fundamental mode, leaving only to obtain the complex amplitude and the propagation constant of the latter.

\subsection{Propagation constant} \label{sec:propagation}

The propagation constant $\beta$ of a single nanocylindrical structure can be obtained by solving the following dispersion relation \cite{Yeh2008, cheng1989field}: 
\begin{align}
     \left(\frac{\beta n}{a}\right)^2 \left( \frac{1}{p_1^2} + \frac{1}{q_2^2} \right)^2  &= \left[ \frac{J'_n(p_1 a)}{p_1 J_n(p_1 a)} + \frac{K'_n(q_2 a)}{q_2 K_n(q_2 a)} \right]  \notag \\
     & \qquad \cdot \left[ \frac{k_1^2 J'_n(p_1 a)}{p_1 J_n(p_1 a)} + \frac{k_2^2 K_n'(q_2 a)}{q_2 K_n(q_2 a) } \right] \label{eq:dispersion}
\end{align}
with $p_1^2 = \omega^2 \mu_0 \epsilon_1 - \beta^2$, $q_2^2 = \beta^2 - \omega^2 \mu_0 \epsilon_2 $. $\omega$, $\mu_0$, $\epsilon_1$ and $\epsilon_2$ are the common physical constants. $J_n$ and $K_n$ are the Bessel function of the first kind and the modified Bessel function of the second kind order $n$ respectfully with prime as the derivative of those function.

From the propagation constant of an isolated structure, it is necessary to add a correction to consider the coupling effects, knowing that a metasurface is composed of an array of equidistant nanostructures. To do so, Coupled Mode Theory (CMT) is often used because of its simplicity, but multiple approximations should be met for its computation to be accurate \cite{Eyges1981}. To consider the periodicity, we will apply a correction to the propagation constant following the Minot method \cite{Minot2010}. Using the slowly-varying envelope and assuming a coupling $C$ which is much smaller than the propagation constant, a long-established approximation of the effective propagation constant of an array of waveguides is expressed as
\begin{align}
    \beta_{\text{array}} = \beta_{\text{isolated}} + \frac{4 \pi}{\lambda} C \cos(k_x S) ,
\end{align}
where $S$ the distance between waveguides, $k_x$ is the transverse Floquet-Bloch waves and $C$ the coupling. Here, we will assume that $k_x = \frac{p+1}{N+1} \pi$ for the $p$ structure within an $N$-size array \cite{Minot2010}. The coupling $C$ can be computed following
\begin{align}
    C = \frac{k_0^2}{2 \beta_{\text{isolated}}} \left( n^2-1 \right) \frac{\langle001 \rangle}{\langle00\rangle} ,
\end{align}
with $\langle00\rangle$ and $\langle001\rangle$ expressed as
\begin{align}
   \langle m \ m' \rangle &= \iint\limits_{\text{whole array}} M_m^* M_{m'} dA ,\\
    \langle m \ m' \ n \rangle &= \iint\limits_{\text{n section cylinder}} M_m^* M_{m'} dA  ,
\end{align}
where $M_p$ correspond to the transverse field distribution of the $p$ structure of the array. For our implementation, $N=2$ and $p=0$ are used, corresponding to the correction of the closest neighbour. For more details on the calculations applied for cylinder structures, see appendix A.

\subsection{Complex amplitude}

Using the Snell-Descartes law, one obtains the transmission angles at the first and second sides of the nanocylinder. Because of the high contrast of the refractive indices often used in dielectric metasurfaces \cite{Chang2018, Kamali2018}, multiple scattering events might occur. To take this phenomenon into account, we consider the Fabry-Perot effect of the nanostructure by introducing a simple geometric sum formulated as 
\begin{align}
    \tau = \frac{t_1 t_2}{1-r_1 r_2 e^{-i 2 \beta H}}, 
\end{align}
where $t_{1,2}$, $r_{1,2}$ correspond to the transmission and reflection coefficients at the first or second interface of the nanofin, and $\beta$ is the propagation constant inside of the nanocylinder of height $H$.

Finally, from Eq. \ref{eq:phase_general}, the output phase-shift $\varphi$ produced by a nanorod of radius $a$ considering only the fundamental mode is given by
\begin{align}
    \varphi(a) = \arg \qty{ \frac{t_1 t_2 e^{-i\beta_{\text{array}}(a) H} }{1-r_1 r_2 e^{-i 2 \beta_{\text{array}}(a) H}}}.
\end{align}

\section{Meta-atoms properties} \label{sec:meta_proper}

\begin{figure}[h]
    \centering
    \begin{subfigure}{0.49\linewidth}
        \centering
        \includegraphics[width=0.95\linewidth]{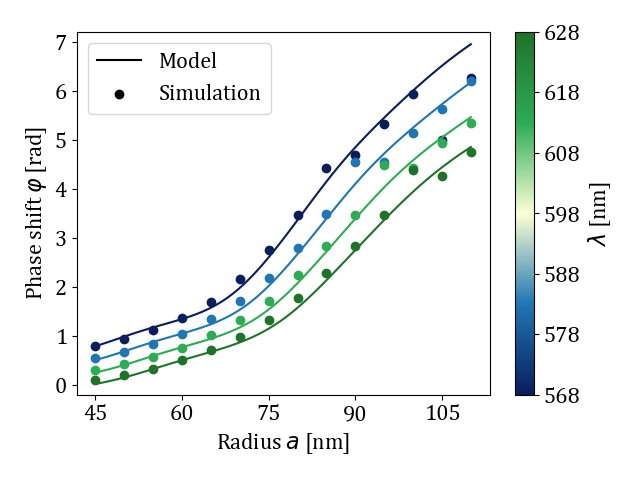}
        \caption{On-axis illumination}
        \label{fig:wave_depend2_fov0}
    \end{subfigure}
    \begin{subfigure}{0.49\linewidth}
        \centering
        \includegraphics[width=0.95\linewidth]{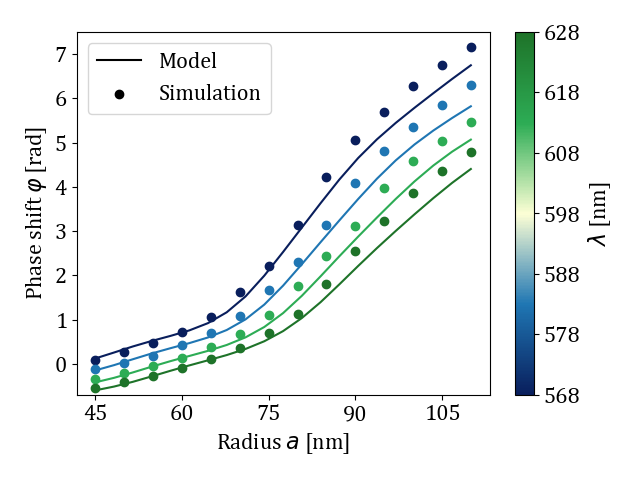}
        \caption{Illumination at $20^\circ$ field of view}
        \label{fig:wave_depend2_fov20}
    \end{subfigure}
    \caption{Output phase shift at various wavelengths for our model and Lumerical simulations. TiO$_2$ meta-atoms are 600 nm high on a SiO$_2$ substrate with a spacing of 350 nm between structures (n$_{\text{TiO}_2}$=2.6139 and n$_{sub}$=1.4584 at $\lambda = 588$ nm) with TM plane wave illumination beam.}
    \label{fig:wave_depend2}
\end{figure}

The semi-analytical model allows for computing the phase shift introduced by those nanostructures under various illuminations. Fig. \ref{fig:wave_depend2_fov0} presents the phase shift $\varphi$ for 600 nm high TiO$_2$ nanocylinders on a SiO$_2$ substrate for wavelengths $\lambda$ from 568 nm to 628 nm. The structures at a varying radius between 45 and 110 nm are located 350 nm apart. These parameters were used since they correspond to metasurfaces often represented in the literature \cite{Liang2018, Decker2019, Shen2020}. As one can observe, the phase can be accurately modelled for various wavelengths when compared to FDTD simulations using Lumerical software \cite{Lumerical}. We can also extend our analysis to illuminations at various incidence angles, as shown in Fig. \ref{fig:wave_depend2_fov20} where the incident beam of light comes at a $20^\circ$ field of view. 

To assess the magnitude of the error in our predictions relative to numerical simulations and the impact this would have in a lens design context, we refer to the work of Patoux et al. \cite{Patoux2021ChallengesIN}. Through their research regarding nanofabrication errors and their effect on metasurfaces performances, they concluded that the fabrication error must be limited to 10 nm for a metalens to maintain acceptable quality. We have identified that the deviations observed in Fig. \ref{fig:wave_depend2} are equivalent to a variation of 1-4 nm in radius on average which is well below the specified limit of 10 nm. Furthermore, a numerical study on the variation of the Strehl ratio showed that errors similar to those shown in Fig. \ref{fig:wave_depend2} still lead to a metalens with a Strehl ratio above 0.9. Such results confirm that the accuracy of our model compared to FDTD simulations is sufficient to represent metasurfaces with performance limited by nanofabrication errors. Therefore, our approach is relevant to computing the properties of the nanostructures and shows the interest of a semi-analytical model to perform a quick initial and optimized design.

\section{Optical design with metasurfaces} \label{sec:design+perfo}

To effectively integrate metasurfaces into the design process, it is of utmost importance that the designer can modify and optimize its characteristics just like any other optical component of the system. Therefore, the model's implementation in the design process was carried out with this objective as a priority. In this section, a design methodology is proposed and then applied to present "meta-refractive" hybrid optical systems as a demonstration.

\subsection{Design process}

\begin{figure*}[h]
    \centering
    \includegraphics[width=0.9\textwidth]{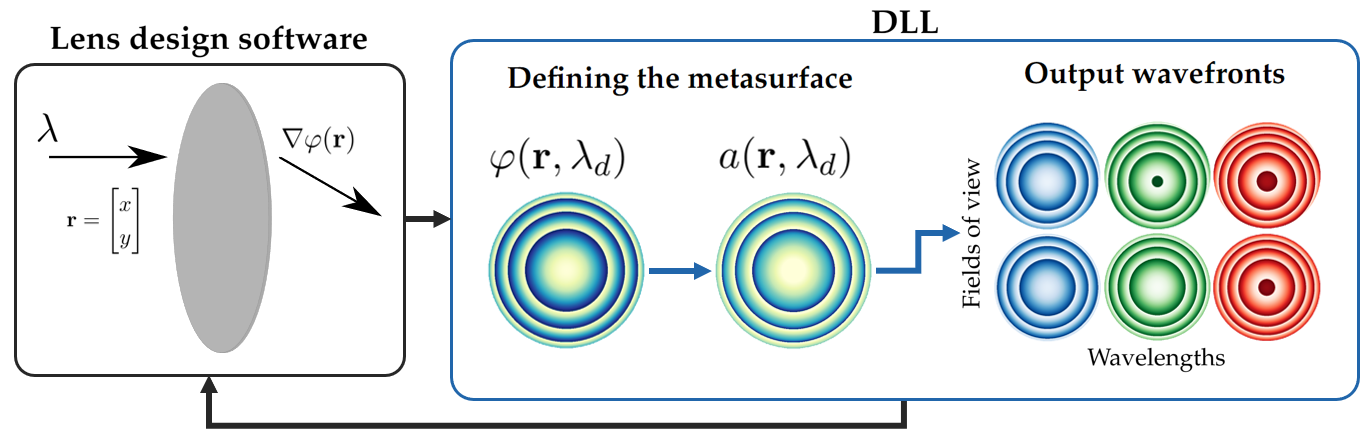}
    \caption{Flowchart of the ray-tracing implementation in Zemax using a custom DLL to compute the nanocylinder's distribution, the output wavefront and the incident rays' refraction angle. }
    \label{fig:Flowchart}
\end{figure*}

The optical design software Zemax has been selected for our first proof of concept. Its ability to process user-defined surfaces via a dynamic link library (DLL) is an adequate solution to handle calculations related to the semi-analytical model. Its operating principle is shown schematically in Fig. \ref{fig:Flowchart}. First, Zemax initiates rays and propagates them until the metasurface is reached. Once intercepted, the DLL starts by defining the metasurface from the characteristics of the meta-atoms and the target phase shift profile (subsection \ref{sec:process1}). The model then computes the wavefronts so that the refraction angles of the rays are given as output by the DLL (subsection \ref{sec:process2}). Exiting the DLL, traditional ray-tracing is resumed across the rest of the optical system. Ultimately, the lens designer can use performance analysis tools to evaluate the system (subsection \ref{sec:process3}).

\subsubsection{Defining the metasurface} \label{sec:process1}
Directly from the Zemax GUI, a lens designer can choose suitable materials and dimensions for the nanostructures. The first step of the design process starts by defining the physical characteristics of the structures to be used: material, dimensions and spacing between structures. For a given height, it is necessary to define an interval of radius $(a_{min}, a_{max})$ allowing an adequate phase covering, i.e. $\abs{\varphi(a_{max}) - \varphi(a_{min})} \geq 2\pi$. A lens designer can find such information with simple preliminary analyzes using the model corroborated by numerical simulations. 

The design phase shift can be expressed in two ways. First, if one desires to produce metalenses specifically designed for focusing, a hyperbolic phase profile is best suited (Eq. \ref{eq:hyperbolic_phase}). Such phase shift leads to a perfect focal spot on the optical axis at a distance $f$ for a plane incident wavefront. However, for a metasurface located between multiple elements, a polynomial phase profile offers much-needed additional degrees of freedom. The phase relation should be defined as Eq. \ref{eq:polynomial_phase}.
\begin{align}
    \varphi_1(r) &= \frac{2\pi}{\lambda} (f - \sqrt{r^2+ f^2}) \label{eq:hyperbolic_phase}, \\
    \varphi_2(r) &= \sum_i a_{i} \qty(\frac{r}{R})^{2i} \label{eq:polynomial_phase}.
\end{align}

\subsubsection{Computing the output wavefront} \label{sec:process2}

Once the metasurface target phase shift is defined, the model is then put to work. Starting from the design phase profile $\varphi(\vb{r}, \lambda_d)$, the nanorods' distribution $a(\vb{r}, \lambda_d)$ that reproduces exactly $\varphi(\vb{r})$ at the design wavelength $\lambda_d$ is found. To do so, the phase relation developed in section \ref{sec:theory} should be inverted in order to obtain a distribution of propagation constant as $\beta = \varphi(r) /H $. Since the only unknown of this relation is now the radius $a$ of the nanocylinder, the radius distribution needed to obtain a target phase distribution is computed. 

For the sake of this numerical application, a continuous distribution of structures of varying dimensions is assumed (see inset of Fig. \ref{fig:Flowchart}). From the spatial distribution of the structures, the model is again used to determine the resulting wavefront for each wavelength and field of view. Finally, the ray-tracing equations developed previously (Eq. \ref{eq:cosine_l} to \ref{eq:cosine_n}) define the angles of refraction of the incident beam from the phase gradient of the metasurface. Finally, the ray's updated information is given back to Zemax so that ray tracing can continue throughout the system.

\subsubsection{Coming up with a design} \label{sec:process3}
One of the advantages of our approach is to be able to exploit directly the optimization algorithms included in the design software. Thus, it is possible to optimize for the most efficient optical system at several wavelengths and fields of view, with a proper merit function, while considering the unique behaviour of metasurfaces. Moreover, a lens designer can study how a slight variation in the dimensions of the structures can impact the performance and may enhance the system. 

The optimization completed and the performance deemed sufficient, one might want to perform a confirmation with full-wave simulations. The wavefront obtained by simulations, possibly slightly different from the one predicted by the model, can be substituted in Zemax$^\textrm{\textregistered}$. If necessary, fine adjustments can be made to downstream elements to compensate for these differences and regain the target optical performances. This way, optimizing performances using the model in the first place provides an adequately accurate starting point that only requires realistic adjustments when moving to numerical FDTD results.

Ultimately, the resulting metasurface is fabricated using conventional nanofabrication methods. Specialized software such as MetaOptics \cite{Dharmavarapu:20} makes it possible to obtain a technical drawing of the distribution of nanostructures to be designed. 

\subsection{Hybrid refractive and metalens designs}

\begin{figure*}[h]
    \centering
    \includegraphics[width=0.98\textwidth]{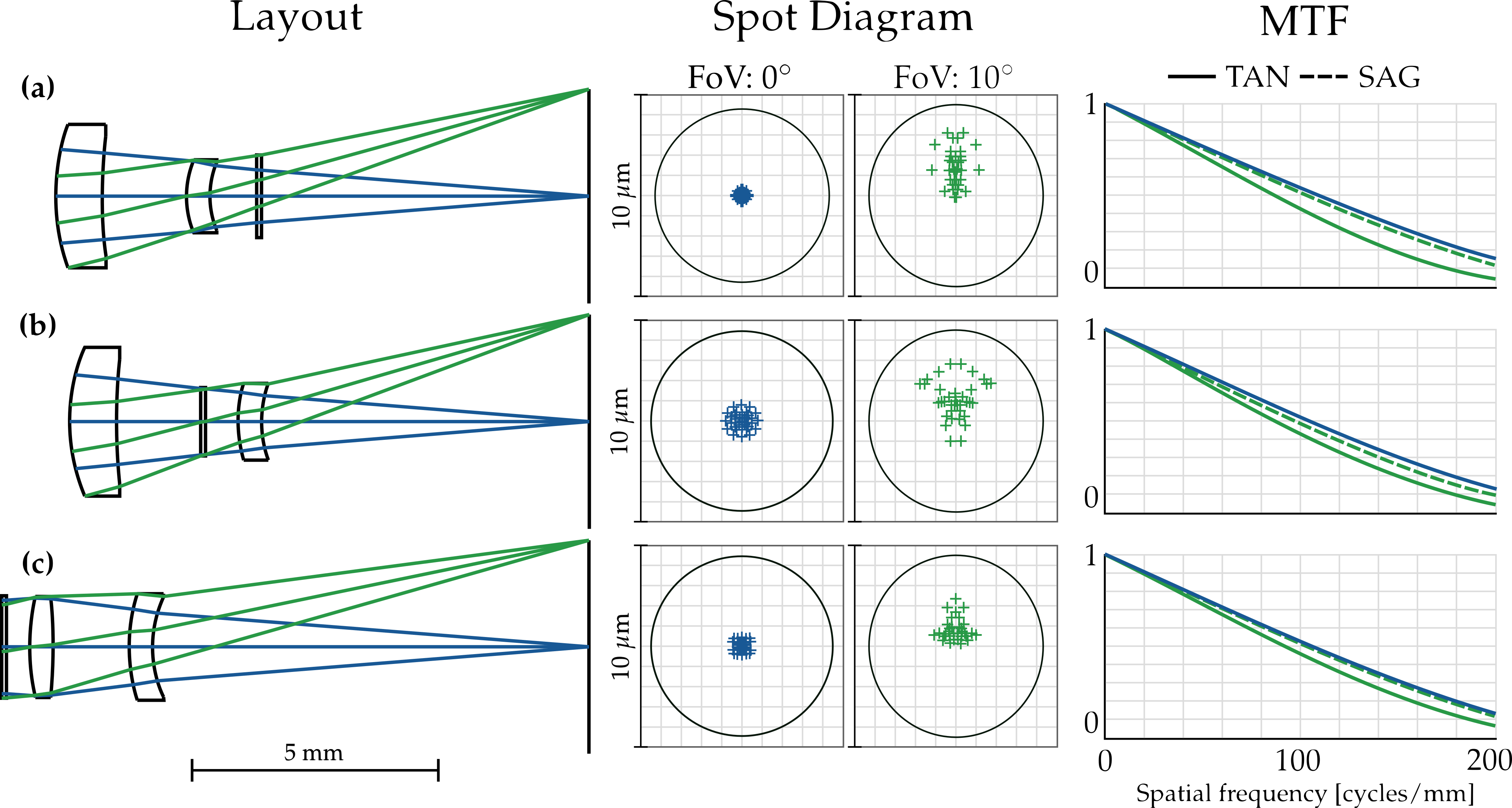}
    \caption{Variations of a $f/\# = 6$ triplet using two refractive lenses and a metasurface ($\lambda = 588$ nm, FoV = $\pm 10^\circ$). The metasurface is either the last \textbf{(a)}, the second \textbf{(b)} or the first element \textbf{(c)}. For each configuration, the spot diagrams and the modulation transfer function (MTF) were chosen as a criterion for performance evaluation.}
    \label{fig:Designs}
\end{figure*}

To demonstrate the efficiency and versatility of our method for obtaining metasurface-based optical systems, we now present several variations of a triplet composed of two refractive elements and a metasurface (see Fig. \ref{fig:Designs}). The decision to design a triplet was taken considering that this type of system is the simplest that can correct all of the seven Seidel aberrations over a wide field of view \cite{Kidger2001}.

According to our process, the first step is defining the structures used in the design. For continuity purposes, we have chosen the nanocylinders presented in section \ref{sec:meta_proper} for the metasurface, knowing that these respect the phase coverage condition and that the model agrees with numerical FDTD simulations. We then arranged the metasurface as either the last (Fig. \ref{fig:Designs}a), second (Fig. \ref{fig:Designs}b) or first (Fig. \ref{fig:Designs}c) element among a combination of two lenses. The arbitrary goal being to design an imaging system with an $f/\# = 6$ suitable for a $\pm 10^\circ$ field of view (FoV), we optimized the metasurface's phase shift profile in addition to the characteristics of the other lenses to obtain a system with almost diffraction-limited performance. For reasons mentioned earlier, we chose a polynomial phase profile over its hyperbolic counterpart to achieve this task.

All three systems display good optical performances, a fact suggested by the focal spots of a smaller size than 10 microns. The Airy spot being 8.6 microns in diameter for these systems means that almost diffraction-limited performance is achieved for the whole field of view. An analysis of the modulation transfer function (MTF) supports these conclusions. Indeed, high contrast is achieved at resolutions suitable for most imaging requirements. As mentioned previously, we can expect a slight variation of the predicted phase compared to the FDTD simulations generating a degradation in the optical quality of the metasurface. However, we have determined that for errors listed in Fig. \ref{fig:Flowchart}, the greatest degradation on the RMS spot size was 0.1 $\mu$m for system \ref{fig:Designs}a), 0.2 $\mu$m for \ref{fig:Designs}b) and 0.4 $\mu$m for \ref{fig:Designs}c). Such variations do not represent a considerable loss of optical quality, highlighting the adequate accuracy of our model.

The optimization process made it possible to adapt the metasurface, just like conventional optics, to the system and thus lead to similar overall performance regardless of its positioning. It demonstrates one of the main advantages of our method: considering metasurfaces as a standard optical element in the optimization process while obtaining an accurate performance evaluation.

\section{Conclusion}

To conclude, we present a new method to implement metasurfaces in the lens design process via a semi-analytical model used in conjunction with ray-tracing software. We had to adapt the ray tracing equations to consider a phase shift induced by metasurfaces in the first place. Instead of going through lengthy numerical simulations to predict the phase shift, the latter is provided by a semi-analytical model that we have developed. Based on an effective medium theory, the Fabry-Perot effect and a closed-form approximation of the coupled-wave theory, our model predicts the phase shift introduced by cylindrical nanostructures. We then proceeded to its implementation in the optical design software Zemax through a DLL. Starting from a target phase shift for the metasurface, the model integrated into the DLL renders the distribution of nanostructures and calculates the output wavefront so that a ray of a given wavelength and orientation can continue through the system. It led to the development of a 3-step feedback process exploiting the ability to include the metasurface in the optimization routine of an optical system. For demonstration purposes, we have designed hybrid refractive and metasurface systems composed of two refractive elements and a metasurface. The latter, being located at distinct positions for each system, was able to adapt just like conventional elements to render a system with near-diffraction-limited performance across the entire field of view. 

Further developments in the generalization of the method to other types of meta-atoms are to be expected. For example, using intrinsically achromatic structures or efficient structures with a large field of view would make it possible to exploit the full potential of metasurfaces in optical design. Ultimately, experimental work could corroborate the accuracy of the semi-analytical model and the relevance of our approach.

\section*{APPENDIX A : CMT for nanocylinders }\label{sec:appendixa}

The sum of mode for the radial component inside $^{(1)}$ and outside $^{(2)}$ of the waveguide can be expressed as \cite{Yeh2008}
\begin{align}
    E_{r}^{(1)}(r,\theta) &= \frac{1}{p_1^2} \sum \left[ - i \beta p_1 A_n J_n'(p_1 r) + \frac{\omega  \mu_0 n}{r} B_n J_n(p_1 r) \right] e^{i n \theta} \\
    E_{r}^{(2)}(r,\theta) &= -\frac{1}{q_2^2} \sum \left[- i \beta q_2 C_n K'_n(q_2 r) + \frac{\omega  \mu_0 n}{r} D_n K_n(q_2 r) \right] e^{i n \theta} ,
\end{align}
where $A_n, B_n, C_n$ and $D_n$ are arbitrary constants. However, to be solutions of the dispersion relation of the waveguide, the constants should follow these relations
\begin{align}
    B_n &= - \frac{   -  \frac{i \omega \epsilon_1 J_n'(p_1 a)}{p_1} + \frac{i \omega \epsilon_2 K_n'(q_2 a)}{q_2}  \cdot \frac{J_n(p_1 a)}{-K_n(q_2 a)}  }{ \frac{\beta J_n(p_1 a)}{a p_1^2} - \frac{\beta K_n(q_2 a)}{a q_2^2} \cdot \frac{J_n(p_1 a)}{-K_n(q_2 a)}  } A_n \end{align} 
or
\begin{align}
 B_n &= - \frac{ \frac{\beta J_n(p_1 a)}{a p_1^2} - \frac{\beta K_n(q_2 a)}{a q_2^2} \cdot \frac{J_n(p_1 a)}{-K_n(q_2 a)} }
   { \frac{i \omega \mu_0 J_n'(p_1 a)}{p1} - \frac{i \omega \mu_0 K_n'(q_2 a)}{q_2} \cdot  \frac{J_n(p_1 a)}{-K_n(q_2 a)} } A_n \\
    C_n &= \frac{J_n(p_1 a)}{K_n(q_2 a)} A_n \\
    D_n &= \frac{J_n(p_1 a)}{K_n(q_2 a)} B_n
\end{align}
where $a$ is the radius of the nanocylinder under consideration. If we only considering the fundamental mode $HE_{11}$, the $\theta$ dependency is not present. Thus, the surface integral in one dimension can be computed with the following equations, where $[f(x)]^*$ is the complex conjugate of the function $f(x)$.
\begin{align}
    \langle00\rangle &= \int_{-\infty}^{-a} \left[E_r(|x|)^{(2)}\right]^* E_r(|x|)^{(2)} dx \notag \\
    & \qquad + \int_{-a}^{a} \left[E_r(|x|)^{(1)}\right]^* E_r(|x|)^{(1)} dx \notag \\ 
    & \qquad \qquad \cdot \int_{a}^{\infty} \left[E_r(|x|)^{(2)}\right]^* E_r(|x|)^{(2)} dx  \\
    \langle001\rangle &= \int_{-a}^{a} \left[E_r(|x|)^{(1)}\right]^* E_r(|x\pm 2\pi|)^{(2)} dx
\end{align}

\section{Funding Information}
Natural Sciences and Engineering Research Council of Canada (NSERC)(RGPIN-2016-05962).

%\section{Disclosures}

\medskip
\noindent\textbf{Disclosures.} The authors declare no conflicts of interest.

% \section{References}

% Bibliography
\bibliography{biblio}

\end{document}